\documentclass[twocolumn,showpacs,preprintnumbers,amsmath,amssymb]{revtex4}

\usepackage{graphicx}
\usepackage{dcolumn}
\usepackage{bm}

\begin{document}


\title{Charge exchange collisions between ultracold fermionic lithium atoms and calcium ions} 

\author{Shinsuke Haze, Ryoichi Saito, Munekazu Fujinaga}
\author{Takashi Mukaiyama}%
\affiliation{Institute for Laser Science, University of Electro-Communications, 1-5-1 Chofugaoka, Chofu, Tokyo 182-8585, Japan}
%
%
%


\begin{abstract}
Charge exchange collisions between ultracold fermionic $^{6}$Li atoms and $^{40}$Ca$^{+}$ ions are observed in the mK temperature range.
The reaction product of the charge exchange collision is identified via mass spectrometry during which the motion of the ions is excited parametrically.
The cross-sections of the charge exchange collisions between $^{6}$Li atoms in the ground state and $^{40}$Ca$^{+}$ ions in the ground and metastable excited states are determined.
Investigation of the inelastic collision characteristics in the atom-ion mixture is an important step toward ultracold chemistry based on ultracold atoms and ions.
  
\end{abstract}
\maketitle

\section{\label{sec:introduction}Introduction}

The chemical properties of particles at low temperatures are expected to be quite different from those at high temperatures. At high temperatures, a chemical reaction rate is determined by the probability of overcoming the energy barrier between the reactant and the product as the result of thermal fluctuations.
However, at low temperatures, where the wave nature of a particle plays an important role in the chemical reaction, tunneling of the energy barrier affects both the reaction and the reaction rate. In such situations, the reaction rate may differ significantly from that predicted by classical mechanics.
In addition, at extremely low temperatures, the quantum statistics of the particles influence the chemical reaction, and either bosonic enhancement \cite{Heinzen} or suppression due to Fermi statistics is expected to be observed.

A hybrid system of ultracold atoms and ions offers an ideal platform for the study of chemical reactions at extremely low temperatures, since sophisticated techniques to efficiently cool and detect atoms and ions have been developed. 
As an example, the elastic and inelastic collision cross-sections between ultracold atoms and ions have been successfully determined for various atom-ion combinations \cite{Grier, Zipkes1, Schmid, Rellergert, Ratschbacher, Chang, Ravi, Haze}.
More recently, a charge exchange collision, which is one of the elementary processes occurring in chemical reactions, was observed at the single-particle level using a trapped $^{174}$Yb$^+$ ion immersed in a Bose-Einstein condensate of $^{87}$Rb atoms \cite{Ratschbacher}.
In fact, it is necessary to work in the ultralow temperature regime, at which only a few partial waves contribute to the atom-ion interactions, in order to experimentally observe the quantum statistical nature of a chemical reaction in an atom-ion hybrid system.

Typically, the threshold energy of the quantum collision regime is given by $E_{\rm{th}}=\hbar^{4}/2C_{4}\mu^{2}$, where $\mu=m_{\rm{i}}m_{\rm{a}}/\left(m_{\rm{i}}+m_{\rm{a}}\right)$ is the reduced mass of an atom of mass $m_{\rm{a}}$ and an ion of mass $m_{\rm{i}}$, and $C_{4}=\alpha Q^{2}/4\pi\epsilon_{0}$, where $\alpha$, $Q$, and $\epsilon_{0}$ are the atomic polarizability, the charge of the ion, and the vacuum permittivity, respectively.
It is evident from the expression for $E_{\rm{th}}$ that it is advantageous to choose an atom-ion combination with a small reduced mass when attempting to work in the quantum collision regime.
In addition, recent theoretical research has demonstrated that the heating effect in an atom-ion hybrid system resulting from ion micromotion (rapidly driven motion in a Paul trap) can be minimized by employing a small $m_{\rm{a}}/m_{\rm{i}}$ ratio \cite{Cetina}.
Knowing this, the use of Li atoms as the neutral species in the atom-ion combinations is helpful during the study of ultracold quantum chemistry.

In this article, we report the observation of charge exchange collisions in a mixture of $^{6}$Li atoms and $^{40}$Ca$^{+}$ ions in the mK temperature regime.
During this work, the loss of ions due to inelastic collisions between the $^{6}$Li atoms in the ground hyperfine state and ions in both the ground state ($S_{1/2}$) and excited states ($D_{3/2}$ and $D_{5/2}$) was observed.
In addition, by identifying the collision products as Li$^{+}$ ions via mass spectrometry with parametric excitation, we were able to ascertain that the inelastic collisions were charge exchange collisions.
Since the reduced mass of the Li-Ca$^{+}$ combination is relatively small compared to that of other recently realized combinations, the present system has a higher $E_{\rm{th}}$ value and a lower heating rate due to micromotion.
The calculated $E_{\rm{th}}$ for the $^{6}$Li-$^{40}$Ca$^{+}$ mixture is $k_{\rm B} \times 10 \mu$K, a value that is much greater than the reported values of $k_{\rm B} \times$ 44 nK for $^{87}$Rb-$^{138}$Ba$^+$ \cite{Schmid}, $k_{\rm B} \times$ 53 nK for $^{87}$Rb-$^{174}$Yb$^+$ \cite{Zipkes1}, and $k_{\rm B} \times$ 280 nK  for $^{40}$Ca-$^{174}$Yb$^+$ \cite{Rellergert}. 
The heating rate calculated for the Li-Ca$^{+}$ combination based on the model by Cetina et al. \cite{Cetina} is 0.5 $\mu$K/s, and is therefore approximately two orders of magnitude smaller than the values reported for other previously researched combinations. This value is also sufficiently low so as to maintain the system in the quantum collision regime over a reasonable experimental timescale.
Although the energy scale of our hybrid system is currently dominated by the ion temperature ($\sim$ mK), which remains above the energy threshold of the quantum collision regime, this particular combination is nevertheless one of the most promising candidates for the study of ultracold atom-ion chemical reaction.

\section{\label{sec:experimental setup}Experimental setup}

\begin{figure}[t]
\includegraphics[width=8cm]{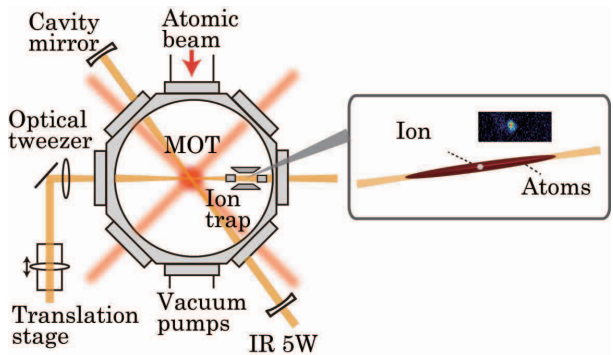}\\
\caption{\label{fig:1} (Color online) Schematic drawing of the experimental setup. A single ion is trapped in a linear Paul trap and immersed in a gas of fermionic atoms in an optical dipole trap. Atoms are transported with optical tweezers following the cavity-enhanced optical dipole trap.} 
\end{figure}

In the present study, Ca$^{+}$ ions and $^6$Li atoms are prepared at different positions in the trap chamber and mixed with one another after all the cooling processes are complete.
Figure \ref{fig:1} shows a schematic of the experimental setup.
Here, Ca$^{+}$ ions are held in a linear Paul trap composed of four RF electrodes and two endcaps located 5 cm from the center of the chamber \cite{Haze}.
The ion trap is driven with a 2.9 MHz RF field and the trap frequencies of the Ca$^{+}$ ions are ($\omega_{\rm{r}}, \omega_{\rm{z}}$) = 2$\pi\times$(170, 35) kHz, where $\omega_{\rm{r}}$ and $\omega_{\rm{z}}$ denote the radial and axial trapping frequencies, respectively.
The radial and axial motions of the ions are Doppler-cooled via the $S_{1/2}$-$P_{1/2}$ transition, using 397 and 866 nm lasers for cooling and repumping, respectively.  
The fluorescent emission from the ions is detected by a photo-multiplier tube (PMT) and a charge-coupled-device (CCD) camera. 
Any stray electric fields are compensated for by applying voltage to the electrodes, such that the ions are placed at the null point of the time-varying RF fields and heating is minimized.
During optimization of the method, the ion motion, synchronized with the RF field, is detected and monitored via an RF-photon correlation method \cite{Berkeland}. 
The ion temperature under optimum conditions is expected to be less than 3 mK, based on estimations using the Doppler broadening of the ion fluorescence spectra during scanning over a range of cooling laser frequencies \cite{Jensen}.

$^6$Li atoms are captured in a magneto-optical trap and subsequently transferred into an optical trap having a relatively large volume and depth ($\sim$1 mK), created with an enhanced light field using an optical resonator \cite{Mosk, Inada}.
These atoms are subsequently transferred into a tightly focused dipole trap and transported to the vicinity of the ion trap electrodes by translational motion of the focusing lens of the trap laser, and mixed with the ions.
These $^6$Li atoms are prepared in the lowest spin states of $S_{1/2}$ ($\mid F=1/2, m_{F}=\pm1/2 \rangle$), with an even population.
To allow for observations of inelastic collisions, evaporative cooling is applied until the atomic temperature is 50 $\mu$K and the number of atoms is $8 \times 10^{4}$.
The radial and axial trapping frequencies of the optical dipole trap are 5.9 kHz and 39 Hz, respectively.

\begin{figure}[t]
\includegraphics[width=8cm]{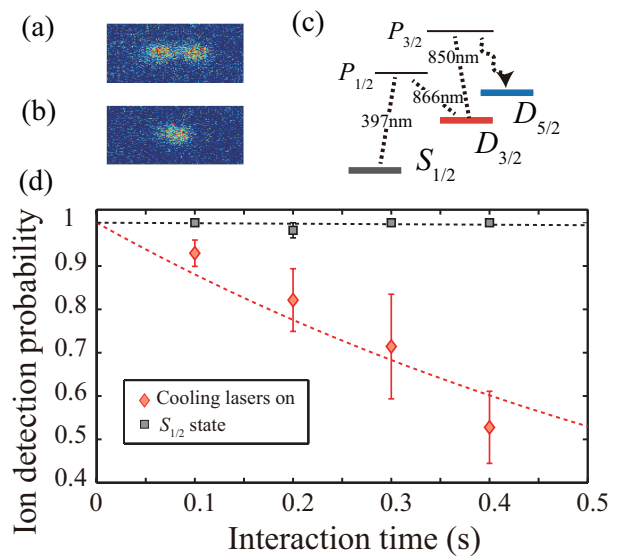}\\
\caption{\label{inelastic} (Color online) Fluorescence images of ions (a) before and (b) after undergoing an inelastic collision. (c) Associated $^{40}$Ca$^{+}$ levels. (d) Relationship between the ion detection probability and the interaction time.}
\end{figure}

\section{\label{sec:results}Results}

In order to observe the inelastic collisions between the atoms and the ions, several to ten ions are initially loaded into the ion trap and spatially overlapped with the atomic cloud. While the ions are mixed with the atoms, the cooling laser at 397 nm and the repumping laser at 866 nm remain on in order to prepare the ions either in the $S_{1/2}$, $P_{1/2}$ or $D_{3/2}$ state, and  the lasers are turned off to prepare the ions to the $S_{1/2}$ state. 
Figures \ref{inelastic}(a) and \ref{inelastic}(b) present fluorescence images of ions before and after undergoing an inelastic collision that causes one ion to completely lose its fluorescence.
Since the ion position is shifted following an inelastic collision, the product of the inelastic collision is lost from the ion trap.
Figure \ref{inelastic}(d) plots the ion detection probability as a function of interaction time. Here the red diamonds indicate data for ions exposed to the cooling and repumping lasers, while the black squares indicate ions prepared in the ground state ($S_{1/2}$) .
In the case of ions exposed to the cooling and repumping lasers, the data show an exponential decay, with an associated decay rate of 1.3(2) s$^{-1}$.
In contrast, ions in the ground state are stable in response to inelastic collisions and the inelastic collision rate is estimated to be less than the 10$^{-2}$ s$^{-1}$, which is the lowest limitation of our measurement.
The effect of elastic collisions between atoms and ions during the interaction time is negligible in the current experiment.
In our experiment, the trap depth for the $^6$Li atoms is way lower than the kinetic energy of the ions and therefore elastic collisions simply remove $^6$Li atoms from the trap.
Within the time scale of our measurement, we observe neither discernible decrease of number of atoms nor a heating of the atomic cloud.

\begin{figure}[t]
\includegraphics[width=6cm]{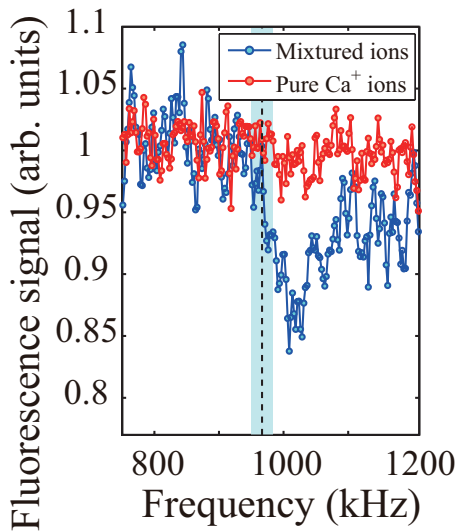}\\
\caption{\label{spectrometry} (Color online) Results from mass spectrometry of trapped ions. The vertical axis shows the normalized fluorescence from the Ca$^{+}$ ions, while the horizontal axis shows the frequency of the applied oscillating field. A reduction in the ion fluorescence indicates the presence of Li$^{+}$ ions in the trap.} 
\end{figure}

During this measurement, the ion's population is distributed to the $S_{1/2}$, $P_{1/2}$ and $D_{3/2}$ states, and it is possible that collisional quenching, where the $D$ state ion is quenched to $S_{1/2}$ state due to inelastic collisions with atoms, may occur. However, the expected recoil energy of the Ca$^{+}$ ions as a result of collisional quenching from the $D_{3/2}$ states is 0.24 eV, which is four times smaller than the trap depth in the current apparatus and therefore we believe that the contribution of this phenomenon to the inelastic loss is small.

The inelastic collisions can result in either charge exchange or molecular ion formation and, in order to determine a detailed mechanism, mass spectrometry \cite{Zipkes2,Schmid,Hall} of the reaction products is performed following the inelastic collisions.
In these trials, a pure Ca$^{+}$ ion cloud consisting of $\sim$400 ions is mixed with Li atoms for 4 s. 
Subsequently, an oscillating electric field is applied to one of the ion trap electrodes to excite the radial motions of the ions.
Fluorescence emission from the Ca$^{+}$ ions is detected by a PMT and the photon counts are recorded while sweeping the frequency of the applied oscillating field.
During these measurements, the amplitude and frequency of the RF field are adjusted such that the q parameters for the possible reaction products (Li$^{+}$ for the charge exchange collisions and CaLi$^{+}$ and Li$_{2}^{+}$ for the molecular ion formation) are maintained in the stable region of the ion trap \cite{Leibfried}. (c.f. q = 0.50 for Li$^{+}$.)

The experimental mass spectrometry data are shown in Fig. \ref{spectrometry}.
Here the blue and red dots indicate the fluorescence from the ions before (red) and after (blue) mixing with the atoms, as functions of the frequency of the applied oscillating field.
The blue dots show that the reduction in the ion fluorescence peaked at 1020 kHz, which is an indication of the resonant motional excitation of the Li$^{+}$ ions.
The vertical dashed line at 966 kHz indicates the expected resonance frequency of the Li$^{+}$ ions, as calculated from the measurement of the resonant radial frequency of the Ca$^{+}$ ions observed at 145 kHz.
The shaded vertical band indicates the uncertainty of the resonance peak, based on the spectral broadening of the measured resonance signal of the Ca$^{+}$ ions.
We believe that the slight discrepancy between the expected and experimental resonant frequency values of 5 $\%$ results from a Coulomb interaction-induced shift in the mass spectra of multispecies ion crystals \cite{Roth}.
The same measurements are also performed around the resonant frequencies of the CaLi$^{+}$ and Li$_{2}^{+}$ molecular ions, although no resonant excitation signals are observed.
From these results, it is evident that the observed inelastic loss is predominantly due to charge exchange collisions.

\begin{figure}[t]
\includegraphics[width=8cm]{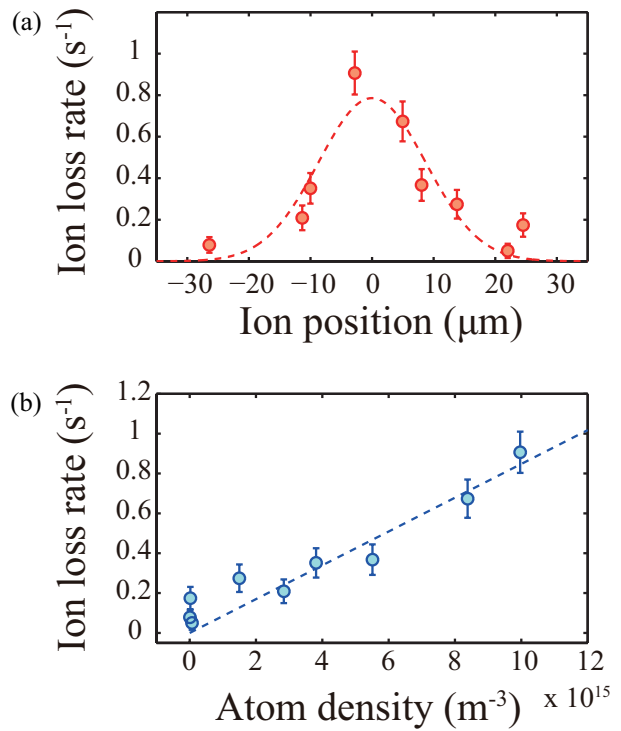}\\
\caption{\label{profile} (Color online) Ion loss rates as functions of (a) ion position, in which the dashed curve shows the atomic density profile calculated from the atomic temperature and trapping frequencies, and (b) atomic density, in which the dashed line shows the linearity of the relationship.}
\end{figure}

We subsequently determined the atomic density dependence of the ion loss rate.
In these trials, ions are moved inside an atomic cloud and the atomic density at which the ions overlapped is controlled.
Here, the position of ions in the radial direction is varied by applying DC voltage to one of the ion trap electrodes.
The resulting data showing the relationships between the measured loss rate and the atom-ion displacement in the radial direction are plotted in Fig. \ref{profile}(a).
By fitting the data with a Gaussian function, a value of 10(2) $\mu$m is determined for the 1/e width.
This value is consistent with the atomic cloud size of 12.3 $\mu$m estimated from the measured atomic temperature and trapping frequencies.
The expected spatial extent of the ions is 1.4 $\mu$m at the center of the atomic cloud.
Although the ions are slightly displaced from the null point of the RF potential when changing the atomic density, the spatial extent of the ions is estimated to be less than 3.2 $\mu$m (corresponding to 16 mK). 
The collision energy is increased as the ion is displaced from the null point, whereas the inelastic loss rate should be independent of the collision energy in a classical collision regime, as predicted \cite{Langevin}. 
This behavior has been previously demonstrated during atom-ion inelastic collisions in a Yb$^{+}$-Rb mixture \cite{Zipkes2}.
Figure \ref{profile}(b) shows the ion loss rate as a function of the atomic density.
Here we find that the ion loss rate linearly increases as the atomic density increases, suggesting that the inelastic collisions observed in this experiment are atom-ion two-body processes. This is in good agreement with the data in Fig. \ref{spectrometry}, which shows that the inelastic collisions represent charge exchange processes.

\begin{figure}[t]
\includegraphics[width=8cm]{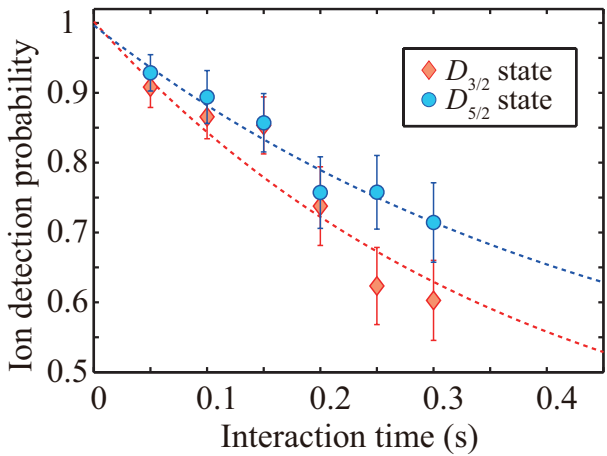}\\
\caption{\label{Dstate} (Color online) Ion detection probabilities as functions of interaction time for ions in the $D_{3/2}$ and $D_{5/2}$ states. The dashed curves are the fitting results with the solution to the rate equations discussed in the text.}
\end{figure}

To further investigate the characteristics of these inelastic collisions, we determined the rates of the charge exchange collisions of ions in the metastable $D$ states. In these trials, the ions are pumped to either the $D_{3/2}$ ($D_{5/2}$) state just prior to overlapping them with atoms
by irradiation with the 397 nm (397 nm and 850 nm) lasers for 20 msec following Doppler cooling (see the energy diagram shown in Fig. \ref{inelastic}(c)).
Figure \ref{Dstate} shows the ion detection probability as a function of interaction time for ions in the $D_{3/2}$ (red diamonds) and $D_{5/2}$ (blue circles) states. These data represent averages over 40 to 100 measurements.
During the atom-ion interaction, the ion experiences heating due to the absence of the cooling lasers, and the ion temperature after a 0.3 s interaction time is estimated to be several hundreds of mK from a separate experiment.
However, we believe that the ion loss rate is not greatly affected by this heating because of the energy independent feature of the inelastic collisions in this energy range.

The ion detection probabilities in the $D$ states can be analyzed using rate equations. Defining the probabilities of finding the ions in the $S$, $D$, and final states after the two-body inelastic collision as $p_{S}$, $p_{D}$, and $p_{\rm inel}$, respectively, the rate equations can be written as $\dot{p}_{D}=-\Gamma p_{D}-\gamma p_{D}$, $\dot{p}_{S}=\gamma p_{D}$, and $\dot{p}_{\rm inel}=\Gamma p_{D}$. In this case, the condition $p_{S}+p_{D}+p_{\rm inel}=1$ must be satisfied, where $\gamma$ is the radiative decay rate of the $D$ state and $\Gamma$ is the inelastic collision rate that is the subject of this study.
In a previous study, the spontaneous decay rates $\gamma$ was found to be 0.86 and 0.85 s$^{-1}$ for the $D_{5/2}$ and $D_{3/2}$ states \cite{Kreuter}.
The dashed curves shown in Fig. \ref{Dstate} are the fitting results obtained using the solution of the rate equations with $\Gamma$ as a fitting parameter.
Here, the initial conditions of the rate equation are $p_{D}(0)=1$, $p_{S}(0)=0$, and $p_{\rm inel}(0)=0$. 
The $\Gamma$ values obtained from the fitting are ($\Gamma_{D_{3/2}}, \Gamma_{D_{5/2}}$) = (1.79(14), 1.30(5)) s$^{-1}$.
The corresponding loss coefficient $K=\frac{\Gamma}{n_{\rm{a}}}$ can be determined from the atomic density $n_{\rm a}$ and these coefficients are found to be ($K_{D_{3/2}}$, $K_{D_{5/2}}$)=($8.2(6)\times 10^{-17}$, $5.9(2)\times 10^{-17}$) m$^{3}$/s.
$K_{S_{1/2}}$ is also estimated from the data shown in Fig. \ref{inelastic} and is determined to be less than $7 \times 10^{-19}$ m$^{3}$/s, a value that is two orders of magnitude below the values for the metastable $D$ states.
Since $K_{D_{3/2}}$ and $K_{D_{5/2}}$ are both less than the Langevin collision coefficient $K_{\rm{L}}=5.0\times10^{-15}$ m$^{3}$/s \cite{Langevin} by roughly two orders of magnitude, and $K_{S_{1/2}}$ is smaller by four orders of magnitude, these inelastic collisions essentially represent an off-resonant process, in clear contrast to the case of homonuclear combinations \cite{Grier}.

Once $K_{D_{3/2}}$ is determined, we are able to calculate the value of $K_{P_{1/2}}$ by analyzing the data shown in Fig. \ref{inelastic}.
Since the cooling and repumping lasers are incident during the interval over which the data are acquired, it is necessary to take into account the fact that the ion population is distributed to the $S_{1/2}$, $P_{1/2}$, and $D_{3/2}$ states.
From the decay of the ion detection probability, the inelastic collision coefficient is calculated to be $K_{SPD}$=8.5(7)$\times$10$^{-17}$.
$K_{SPD}$ can be expressed as $K_{SPD}$=$P_{S}K_{S}$ + $P_{P}K_{P}$ + $P_{D}K_{D_{3/2}}$, where $P_{S}$, $P_{P}$, and $P_{D}$ are the population probabilities in the $S_{1/2}$, $P_{1/2}$, and $D_{3/2}$ states, respectively. 
These population probabilities can, in turn, be estimated based on the experimental conditions, giving values of ($P_{S}$, $P_{P}$, $P_{D}$)=(0.34, 0.14, 0.52).
The loss coefficient for the $P_{1/2}$ state can be obtained from these values and is found to be  $K_{P}$ = 3.0(5)$\times$10$^{-16}$ m$^{3}$/s.

The significant differences between the charge exchange collision coefficients for the $S$ and $D$ states may result from the close lying energy levels Li$^{+}$+Ca (3$d$4$s^{3}D$) of the $D$ states.
Similar results have been reported for experimental trials using the Rb-Yb$^{+}$ system, in which the reaction rates of ions in the metastable $D_{3/2}$ and $F_{7/2}$ states were investigated in detail \cite{Ratschbacher}. Energy level crossing may thus occur in the Li$^{+}$+Ca (3$d$4$s^{3}D$) level and Li + Ca$^{+}$ ($D_{3/2}$ or $D_{5/2}$) levels at sufficiently short distances at which the interaction potentials may be modified when working at high energy levels \cite{Sayfutyarova, Hall}. A quantitative understanding of the charge exchange collision coefficients determined for our system will require further studies of the related energy levels.

\section{\label{sec:conclusion}Conclusion}

In conclusion, we were able to observe inelastic collisions between $^{6}$Li atoms and $^{40}$Ca$^{+}$ ions, and these were confirmed to be charge exchange collisions based on mass spectrometry of the reaction product ions.
These charge exchange collisions were largely suppressed in the case of ions in the electronic ground state, which is advantageous with regard to the sympathetic cooling of Ca$^{+}$ ions when attempting to reach the $\mu$K temperature regime without a significant loss of ions.
In addition, the reaction rate could be dynamically switched by applying a short optical pulse to excite the ions to the metastable states.
This capability of on-off switching can be useful in terms of controlling ultracold chemical reactions and in applying ions as a high-spatial resolution probe for atomic gases in an optical lattice potential.
The next challenge is to cool the atoms and ions below 10 $\mu$K to observe the atom-ion interaction in a quantum collision regime. Atoms can be readily cooled well below the temperature by conventional evaporative cooling, which has already been done in our system. Ions can also be cooled below the temperature by sideband cooling in a typical experimental condition \cite{Monroe}.

\section{\label{sec:acknowledgement}Acknowledgement}

This work was supported by JSPS KAKENHI Grants No.22104503, No. 25800227, and No. 26287090.


\begin{thebibliography}{99}

\bibitem{Heinzen} D. J. Heinzen, R. Wynar, P. D. Drummond and K.V. Kheruntsyan, \prl \textbf{84}, 5029 (2000).

\bibitem{Ravi} K. Ravi, S. Lee, A. Sharma, G. Werth and S. Rangwala, Nature comm. \textbf{3}, 1126 (2012).

\bibitem{Grier} A. Grier, M. Cetina, F. Oru\v{c}evi\'{c}, and V. Vuleti\'{c}, Phys. Rev. Lett. 102, 223201 (2009).

\bibitem{Zipkes1} C. Zipkes, S. Palzer, C. Sias and M. Kohl, Nature \textbf{464} 388 (2010).

\bibitem{Schmid} S. Schmid, A. H\"{a}rter, and J. H. Denschlag, Phys. Rev. Lett. \textbf{105}, 133202 (2010).

\bibitem{Rellergert} W. Rellergert, S. Sullivan, S. Kotochigova, A. Petrov, K. Chen, S. Schowalter and E. Hudson, Phys. Rev. Lett. \textbf{107}, 243201 (2011).

\bibitem{Ratschbacher} L. Ratschbacher, C. Zipkes, C. Sias, and M. K\"{o}hl, Nature Phys. \textbf{8}, 649 (2012).


\bibitem{Chang} Y. -P. Chang, K. Dlugolecki, J. Kupper, D. Rosch, D. Wild, S. Willitsch, Science \textbf{342}, 98 (2013).

\bibitem{Haze} S. Haze, S. Hata, M. Fujinaga, and T. Mukaiyama, Phys. Rev. A \textbf{87}, 052715 (2013).

\bibitem{Cetina} M. Cetina, A. Grier and V. Vuleti\'{c}, Phys. Rev. Lett. \textbf{109}, 253201 (2012).

\bibitem{Berkeland} D. Berkeland, J. Miller, J. Bergquist, W. Itano, and D. Wineland., J. Appl. Phys. 83, 5025 (1998).

\bibitem{Jensen} M. J. Jensen, T. Hasegawa and J. J. Bollinger,  Phys. Rev. A \textbf{70}, 033401 (2004).

\bibitem{Mosk} A. Mosk, S. Jochim, H. Moritz, Th. Elsasser, M. Weidemuller, and R. Grimm, Opt. Lett. \textbf{26}, 1837 (2001).

\bibitem{Inada} Y. Inada, M. Horikoshi, S. Nakajima, M. Kuwata-Gonokami, M. Ueda and T. Mukaiyama, \prl \textbf{101}, 180406 (2008).

\bibitem{Zipkes2} C. Zipkes, S. Palzer, L. Ratschbacher, C. Sias, and M. K\"{o}hl, Phys. Rev. Lett. \textbf{105}, 133201 (2010). 


\bibitem{Hall} F. Hall, M. Aymar, N. Bouloufa-Maafa, O. Dulieu and S. Willitsch, Phys. Rev. Lett. \textbf{107}, 243202 (2011).

\bibitem{Leibfried} D. Leibfried, R. Blatt and D. Wineland, Rev. Mod. Phys. \textbf{75}, 281 (2003). 

\bibitem{Roth} B. Roth, P. Blythe, and S. Schiller, Phys. Rev. A \textbf{75}, 023402 (2007). 

\bibitem{Langevin} P. Langevin, Ann. Chim. Phys. \textbf{5}, 245 (1905).

\bibitem{Kreuter} A. Kreuter, C. Becher, G. P. T. Lancaster, A. B. Mundt, C. Russo, H. Haffner, C. Roos, W. Hansel, F. Schmidt-Kaler, R. Blatt, M. S. Safronova, Phys. Rev. A \textbf{71}, 032504 (2005).


\bibitem{Sayfutyarova} E. Sayfutyarova, A. Buchachenko, S. Yakovleva, and A. Belyaev, Phys. Rev. A \textbf{87}, 052717 (2013). 

\bibitem{Monroe} C. Monroe, D. Meekhof, B. King, S. Jefferts, W. Itano, D. Wineland, and P. Gould, Phys. Rev. Lett. \textbf{75}, 4011 (1995).

\end{thebibliography}
\end{document}